\documentclass[prl,aps,twocolumn,psfig,preprintnumbers,amsmath,amssymb]{revtex4}
\usepackage{graphicx}\usepackage{epsfig}
\usepackage{amsmath}\usepackage{amssymb}\usepackage{amsfonts}
\usepackage{bm}
\usepackage{exscale}
\usepackage{dcolumn}    
\usepackage{color}
\usepackage{soul}
\usepackage{ulem}
\begin{document}
%
%
\title{The isoscalar features of Pygmy Dipole Resonance: a subtle game of symmetry energy}
%
%
\author{V. Baran$^{1,*}$}
\author{T. Isdraila$^{1}$}
\author{M. Colonna$^{2}$}
\author{D.I. Palade$^{1}$}
\author{A.I. Nicolin$^{1,3,\dagger}$}
\affiliation{$^{1}$Faculty of Physics, University of Bucharest, Atomistilor 405, Magurele, Romania}
\affiliation{$^{2}$INFN-LNS, Laboratori Nazionali del Sud, 95123 Catania, Italy}
\affiliation{$^{3}$Institute of Space Science, Atomistilor 409, Magurele, Romania}
\affiliation{$^*$baran@fizica.unibuc.ro}
\affiliation{$^\dagger$alexandru.nicolin@fizica.unibuc.ro}

\begin{abstract}
The vibrational structure of the Pygmy Dipole Resonance (PDR) is investigated within a quantum many-body treatment with extended separable interactions able to encode the dependence of nuclear symmetry energy on density. A new picture of PDR is unveiled in terms of a combined dynamics of the neutron skin and of the core isovector polarization, which determines the isoscalar features of PDR while reproducing the isovector properties of Giant Dipole Resonance. The key role played by the variation with density of the symmetry energy on shaping the low-lying dipole response and its isoscalar-isovector structure is underlined. Our results provide insights for the challenge of clarifying the transition from skin oscillation to a highly bulk collective dynamics.
\end{abstract}
%
%
%
%
%
\maketitle
%
%
%
%
%

As a result of quantum correlations in the presence of particle interaction, many-body systems manifest a rich dynamical behaviour, which extends from elementary, single-particle excitations, to collective, sometimes non-perturbative, states with a coherent participation of many constituents \cite{rin1980,fanRMP1992,lip2008}. 


The nature of the low-lying dipolar response in neutron-rich nuclei, below the Giant Dipole Resonance (GDR), observed as a resonance-like structure, exhausting a few percentages of the Energy Weighted Sum Rule (EWSR) \cite{zilPLB2002,adrPRL2005,wiePRL2009,rosPRL2013,wiePRC2018} and ascribed to Pygmy Dipole Resonance (PDR) states, is still under discussion \cite{braPPNP2019,broPS2019}. The experimental results indicate puzzling properties that spark hot debates, demanding more data on a wider isospin and mass range. This data will allow to identify correlations between the measured features of PDR, such as energy position, EWSR, and polarizabilities, and the properties of nuclei, such as mass, isospin, and size of the neutron skin, or the parameters of the symmetry energy of the equation of state \cite{wiePPNP2011,aumPS2013,savPPNP2013}. To this end, experimental facilities dedicated to nuclear photonics \cite{nedPAN2017}, either in operation or currently under commissioning, such as HI$\gamma$S \cite{welPPNP2009} and ELI \cite{camRRP2016,galRPP2018}, can uncover the detailed structure of the PDR using next-generation gamma-ray sources \cite{ndiPRAB2019}.

 Theoretically, the open questions are mainly concerned with the collectivity of these states, the energy centroid position, the fragmentation of the response and the role of symmetry energy \cite{paaRPP2007,paaJPG2010,rocPPNP2018}. In fact, while it was concluded that the symmetry energy term that defines the isospin contribution in the energy per particle
$\displaystyle E/A(\rho,I)=E/A(\rho,I=0)+ b_{sym} (\rho) I^2$ \cite{barPR2005} influences the dynamics of this mode \cite{carPRC2010,barPRC2012,barPRC2013}, an isoscalar component in the vibrational structure was also evidenced \cite{lanPRC2011,rocPRC2012,papPRC2015,zhePRC2016}.

In this Letter we demonstrate in a transparent manner the link between the symmetry energy and the isoscalar nature of PDR, a problem insufficiently clarified until now. We place special emphasis on the role of density dependence of the symmetry energy on the development of the low-lying $E1$ response in neutron rich nuclei. 

Our study is based on a quantum many-body model which relies on the Tamm-Dancoff Approximation (TDA) and Random Phase Approximation (RPA) with extended separable interactions \cite{barPRC2015}. We generalise the Brown-Bolsterli model \cite{broPRL1959,broNPA1961} by introducing a density-dependent particle-hole residual interaction that encodes the behaviour of the symmetry energy below saturation. This model predicts the appearance of two non-trivial states, associated with PDR and GDR, which share all $E1$ EWSR, thereby manifesting specific collective features \cite{barPRC2015}. To address the important questions concerning
the isoscalar-isovector structure of dipolar responses \cite{savPRL2006,tsoPRC2008,paaPRL2009,burPRC2019} and the macroscopic picture of the vibration of nucleons  \cite{ryePRL2002, schPRC2008,lanPRC2015}, it is enough to focus on the simpler TDA approach, since the inclusion of ground state correlations does not modify the main conclusions, see \cite{barPRC2015} for a detailed RPA treatment. 
For the $E1$ excitations, the separable ansatz for the residual interaction is $\displaystyle \bar{V}_{ph',hp'}=\lambda_1D_{ph}D^*_{p'h'}$, where 
$\displaystyle D_{ph}=\langle ph|\hat{D}|0 \rangle= \langle p |\hat{d}| h \rangle \equiv Q_i$ are the particle-hole matrix elements of single-particle dipole operator $\hat{\vec{d}}=\frac{ZN}{A}
(\frac{1}{Z}\sum_{i=1}^Z \hat{\vec{r}}_i-\frac{1}{N}\sum_{i=1}^N \hat{\vec{r}}_i)$ and 
$\displaystyle \hat{D}$ is the one-body dipole operator $\hat{D}=\sum_{\mu,\nu} D_{\mu \nu}a^+_{\mu}a_{\nu}$. The self-consistency requirement between the vibrating potential and the induced density variations for the dipolar field, fixes the coupling constant $\displaystyle \lambda_1$ to $ \lambda_1(\rho_0)=\frac{A^2}{NZ} \frac{10 b_{sym}^{(pot)}(\rho_0)}{A R^2}$, with $R=1.2 A^{1/3}$, proportional to the potential symmetry energy $b_{sym}^{(pot)}(\rho_0)$ at saturation \cite{boh1998, barPRC2015}. 
However, the potential symmetry energy decreases with density, and a smaller value for the coupling constant of the particle-hole interaction is expected at lower densities. Because a fraction of the nucleons is localised in a less dense surface region, we relax the condition of a unique coupling constant for all particle-hole interactions. We assume that only for a subsystem of particle-hole pairs, $i.e., i,j \le i_c$, at the saturation density, the interaction is given by $\displaystyle A_{ij}=\lambda_1(\rho_0) Q_i Q_j^{*}$, while for the subsystem at lower density $\displaystyle \rho_e$, $i.e., i,j>i_c$ , the interaction is characterised by a weaker 
strength $A_{ij}=\lambda_3(\rho_e)Q_i Q_j^{*}$. If $\displaystyle i \le i_c$ and $j > i_c$ or $i > i_c$ and $j \le i_c$, {\it i.e.}, for the coupling between the two subsystems, within a region of intermediate density, we consider $\displaystyle A_{ij}=\lambda_2(\rho_i) Q_i Q_j^{*}$. The values of $\lambda_2$ and  $\lambda_3$ will reflect the behaviour of symmetry energy below normal density and therefore $\displaystyle \lambda_1(\rho_0)>\lambda_2(\rho_i) >\lambda_3(\rho_e)$. 
With this extended separable interaction the equations for the amplitudes that define the TDA excitation operators, 
$\Omega_{TDA}^{+ (n)}=\sum_{i \le i_c}X_{i}^{(n)}a_p^{+}a_h+\sum_{i > i_c}X_{i}^{(n)}a_p^{+}a_h$,
derived using the equation-of-motion method \cite{row2010} within the linearisation approximation, are:  
\begin{eqnarray}
\epsilon_i X_i^{(n)}+\lambda_1 Q_i \sum_{j \le i_c}Q_j^{*}X_j^{n}+\lambda_2 Q_i \sum_{j >i_c}Q_j^{*}X_j^{n} =E_n X_i^{(n)} \nonumber\\
\label{amplitx1}  \nonumber \\
\epsilon_i X_i^{(n)}+\lambda_2 Q_i \sum_{j \le i_c}Q_j^{*}X_j^{n}+\lambda_3 Q_i \sum_{j > i_c}Q_j^{*}X_j^{n} =E_n X_i^{(n)}
\label{amplitx2} 
\end{eqnarray}
for $i \le i_c$ and for $i > i_c$, respectively.
The conditions for nontrivial solutions of (\ref{amplitx2}) together with the normalisation $\sum_j|X_j^{(n)}|^2=1$ will determine the 
energy $\displaystyle E_n$ of the state $| n \rangle =\Omega_{TDA}^{+ (n)}|0 \rangle$, as well as the the corresponding amplitudes that define the particle-hole structure of this state.
Insightful analytical expressions are obtained in the degenerate case $\epsilon_i=\epsilon=\hbar \omega_0=41 A^{-1/3}$ MeV \cite{rin1980}, where calculations are heavily simplified without affecting the generic physical properties. By introducing the quantities $\alpha= \sum_{i \le i_c} |Q_i|^2$, $\beta= \sum_{i > i_c} |Q_i|^2$, from the EWSR associated with the $E1$ excitation corresponding to the unperturbed case, we  notice that $m_1=\int ES(E)dE =\hbar \omega_0 (\alpha+\beta)= \hbar^2/2 m\cdot NZ/A $, where $S(E)$ is the strength function (SF). The values for $\displaystyle \alpha$ and $\displaystyle \beta$ are related to the number of protons $\displaystyle Z_c=Z$ and neutrons $\displaystyle N_c$ which belongs to the nuclear core (nucleons at saturation density), $\displaystyle A_c=N_c+Z_c$, and the number of neutrons considered in excess $\displaystyle N_e$ (nucleons at lower density), $N_e+N_c=N$. We then have  $\displaystyle \hbar \omega_0 \alpha= \hbar^2/2 m \cdot N_c Z /A_c $, 
$\displaystyle \hbar \omega_0 \beta=  \hbar^2/ 2 m \cdot N_e Z^2/A A_c $ \cite{barRJP2012,barPRC2012}. 
In the degenerate case the condition for non-trivial solutions of (\ref{amplitx2}) reduces to:
\begin{equation}
(E_n-\epsilon)^2-(\lambda_1 \alpha + \lambda_3 \beta) (E_n-\epsilon) +(\lambda_1\lambda_3-\lambda_2^2)\alpha\beta=0~~
\label{detTDA}
\end{equation}
which yields two distinct energies $\displaystyle E_c^{(1)},E_c^{(2)}$:
\begin{equation}
E_c^{(1,2)}=\epsilon+\frac{(\lambda_1 \alpha + \lambda_3 \beta)}{2}\left(1\pm \sqrt{1-\frac{4(\lambda_1\lambda_3-\lambda_2^2)\alpha\beta}
{(\lambda_1 \alpha + \lambda_3 \beta)^2}}\right).
\label{enTDA1} 
\end{equation}
From the solutions for the amplitudes:
\begin{equation}
X_i^{(k)}=\frac{Q_i}{\sqrt{\alpha+\eta_k^2\beta}}~,i \le i_c~;~ 
X_i^{(k)}=\frac{\eta_k Q_i}{\sqrt{\alpha+\eta_k^2\beta}}~,i > i_c~, \nonumber
\label{ampTDA}
\end{equation}
with $\displaystyle \eta_k=((E_c^{(k)}-\epsilon)-\lambda_1\alpha)/\lambda_2\beta$, $k=1,2$, we observe that 
the transition amplitudes associated to the two states
\begin{eqnarray}
|n_{c,k}\rangle =\Omega_{TDA}^{(k) +}|0\rangle=~~~~~~~~~~~~~~~~~~~~~~~~~~~~~ \nonumber\\
\left(\sum_{(ph) \le i_c}\frac{D_{ph}}{\sqrt{\alpha+\eta_k^2\beta}}a_p^{+}a_h+ 
    \sum_{(ph) > i_c}\frac{\eta_k D_{ph}}{\sqrt{\alpha+\eta_k^2\beta}}a_p^{+}a_h\right)|0\rangle, \nonumber
\label{ncol12}
\end{eqnarray}
are
$\langle n_{c,k}|\hat{D}|0 \rangle=\sum_i Q_i X_i^{(k) *}=(\alpha+\eta_k\beta)/\sqrt{\alpha+\eta_k^2\beta}$.
Then the total unperturbed transition probability will  distribute only among these two states, {\it i.e.},
$|\langle n_{c,1}|\hat{D}|0\rangle|^2+|\langle n_{c,2}|\hat{D}|0\rangle|^2=\alpha+\beta=\sum_i |Q_i|^2$
indicating collective features for both of them. It is obvious
that in the limit $\lambda_1=\lambda_2=\lambda_3$
we return to the standard situation with only one collective state, since the transition amplitude of the state with higher
energy goes to $\alpha+\beta$, while the transition amplitude of the state with energy $\displaystyle \epsilon$ becomes zero. This clearly shows that the decrease with density of the symmetry energy determines the emergence of an additional collective state, exhausting a part of the $E1$ transition amplitude in the TDA and a fraction of the EWSR in the RPA \cite{barPRC2015}. 
Within this model we investigate the macroscopic picture of nucleon vibrations for these states and inquire about the structure
of the degrees of freedom involved in the PDR response following the  approach of Brown \cite{bro1964} for an $E1$ excitation along the $z$-axis. We assume
that in the configuration space the ground state has a Hartree structure, namely 
$\displaystyle \Psi_{GS}(z_1,z_2,...,z_A)=\phi_1(z_1)\phi_2(z_2)...\phi_A(z_A)$.
If ${\hat \rho}(z)= \sum_{l=1}^{Z+N_c+N_e}\delta({\hat z}_l-z)$ is the local density operator, then the 
ground state density distribution is given by
$\rho_0(z)=\langle GS |{\hat \rho}(z) | GS \rangle=
\sum_{l=1}^{A}\int\delta(z_l-z)|\phi_l(z_l)|^2dz_l
=\rho_{0,p}(z)+\rho_{0,n_c}(z)+\rho_{0,n_e}(z)$.
With the absorption of an E1 photon the final states 
$\displaystyle |n_{c,k}(t)\rangle= e^{-iE_c^{(k)}t/\hbar}\Omega_{TDA}^{(k) +}|GS\rangle$,
have in the configuration space, to first order in the amplitudes $\displaystyle C_k$, the structure:  
\begin{eqnarray}
\Psi_{c,k}(z_1,...,z_A)=\phi_1(z_1)...\phi_A(z_A)+C_{k}e^{-i\omega_kt}   ~~~~~~~~~~~~~~~\nonumber\\
\left( \sum_{l \le A_c}\tau_{3l}z_l \phi_1(z_1)...\phi_A(z_A)
+\sum_{l > A_c}\tau_{3l} \eta_k z_l \phi_1(z_1)...\phi_A(z_A) \right)~,~~~\nonumber
\label{GS}
\end{eqnarray}
where $\displaystyle \tau_{3l}=+1$ $(-1)$ for neutrons (protons), $\displaystyle \omega_k=E_k/\hbar$. 
The time-dependent local density for these states is:
\begin{eqnarray}
\rho(z,t)=\langle n_{c,k}(t)|{\hat\rho}(z)|n_{c,k}(t)\rangle=\rho_0(z) +~~~~~~~~~~~~~~~~~~~~~~~~~ \nonumber \\
2C_{k}\cos\omega_kt \int dz_1 dz_2...dz_A \sum_{j=1}^{A}\delta(z_j-z)\times ~~~~~~~~~~~~~~~~~\nonumber\\
(\sum_{q=1}^{Z}z_q|\phi_1(z_1)|^2...|\phi_A(z_A)|^2-\sum_{q=Z+1}^{A_c}z_q|\phi_1(z_1)|^2...|\phi_A(z_A)|^2~~~~~~~~~\nonumber\\
-\sum_{q=A_c+1}^{A}\eta_k z_q|\phi_1(z_1)|^2...|\phi_A(z_A)|^2) ~~~~~~~~~~~~~~~~~~~~~~~\nonumber
\label{densCS}
\end{eqnarray}
Then the density variation $\displaystyle  \delta\rho(z,t)=\rho(z,t)-\rho_0(z)$ has contributions from 
the protons and neutrons of the core, as well as from the excess neutrons
\begin{equation}
\delta\rho(z,t)\propto z( \rho_{0,p}(z)-\rho_{0,n_c}(z)-\eta_k\rho_{0,n_e}(z) ) \cos\omega_kt~,
\label{densCS}
\end{equation}
showing a Goldhaber-Teller dynamics for these three nucleonic systems with a frequency $\displaystyle \omega_k$ specific to each normal mode. 
This generalises the semiclassical picture of the GDR, described by Brown \cite{bro1964}, through the inclusion of the degrees of freedom associated with excess neutrons. 
For a more detailed analytical analysis of the structure of these vibrations we adopt an approximate
expression for the two energies, valid if $\displaystyle \lambda_1 \alpha\gg\lambda_3 \beta$:  
\begin{equation}
E_c^{(1)}\approx \epsilon+(\lambda_1 \alpha + \lambda_3 \beta)~; 
E_c^{(2)}\approx \epsilon+\frac{(\lambda_1\lambda_3-\lambda_2^2)\alpha\beta}
{(\lambda_1 \alpha + \lambda_3 \beta)}~. 
\end{equation}
The energy  $E_c^{(1)}=E_c^{(GDR)}$ corresponds to the GDR and
\begin{equation}
\eta_1=\frac{E_c^{(GDR)}-\epsilon-\lambda_1\alpha}{\lambda_2\beta}=\frac{\lambda_3}{\lambda_2}>0.
\label{eta1app}
\end{equation}
From (\ref{densCS}) the proton and neutron density oscillations are:
\begin{equation}
\delta\rho_p(z,t)\propto z\rho_{0,p}(z)\cos\omega_{GDR}t~,
\label{densGDRp}
\end{equation}
\begin{equation}
\delta\rho_n(z,t) \propto -z (\rho_{0,n_c}(z)+\frac{\lambda_3}{\lambda_2}\rho_{0,n_e}(z)) \cos\omega_{GDR}t~.
\label{densGDRn}
\end{equation} 
Since $\eta_1$ is positive, all neutrons oscillate against protons, in agreement with the isovector character of GDR. The neutronic density variation results from in-phase oscillations of the neutrons of the core and of excess neutrons, the two terms having the same sign in
(\ref{densGDRn}).

A different picture holds true for the second state, associated to PDR,
which has an energy closer to the distance between shells, $\displaystyle E_c^{(2)}=E_{c}^{(PDR)}\approxeq \epsilon$. In this case
\begin{equation}
\eta_2=\frac{E_c^{(PDR)}-\epsilon-\lambda_1\alpha}{\lambda_2\beta}\approx-\frac{\lambda_1 \alpha}{\lambda_2\beta}<0~.
\label{eta2app}
\end{equation}
The proton and neutron density variations are
\begin{equation}
\delta\rho_p(z,t) \propto z\rho_{0,p}(z)\cos\omega_{PDR}t~,
\label{densPDRp}
\end{equation}
\begin{equation}
\delta\rho_n(z,t) \propto -z(\rho_{0,n_c}(z)+\eta_2\rho_{0,n_e}(z))\cos\omega_{PDR}t~.
\label{densPDRn}
\end{equation}
Since $\eta_2<0$ the excess neutrons oscillate out of phase with the neutrons of the core, a feature determined by the hierarchy among the values of coupling constants, and, moreover, will oscillate in phase with the protons.
\begin{figure}
\begin{center}
\includegraphics*[scale=1.8]{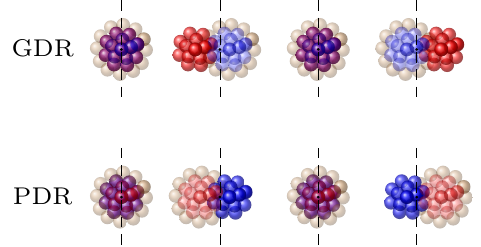}
\end{center}
\caption{(Color online) The macroscopic picture of the protons (red), of neutrons of the core (blue) and of neutrons in excess (grey) vibrations
for the GDR and PDR modes as predicted by the schematic model.}
\label{vibr_gdr_pdr}
\end{figure} 
From Eqs. (\ref{densPDRp}), (\ref{densPDRn}) we observe that the core does not remain inert, but performs an isovector oscillation
with the frequency $\omega_{PDR}$ as illustrated in Fig. \ref{vibr_gdr_pdr}. The core polarization shown by our model represents a significant correction of the classical scenario which interprets the PDR as an oscillation of neutron skin against an inert core \cite{lanPRC2015}.
Assuming $\rho_{0,n_e}/\rho_{0,n_c}\approx N_e/N_c$ the density variation of the neutrons becomes
\begin{equation}
\delta\rho_n(z,t) \propto z\rho_{0,n_c}(z)\left(\frac{\lambda_1 A}{\lambda_2 Z}-1\right)\cos\omega_{PDR}t~.
\label{densPDRnf}
\end{equation}
For $\displaystyle \lambda_1 A>\lambda_2 Z$ an inspection of Eqs. (\ref{densPDRp}), (\ref{densPDRnf}) indicates that
the proton and neutron densities oscillate in phase, revealing an isoscalar character of the PDR. 
\begin{figure}
\begin{center}
\includegraphics*[scale=0.30]{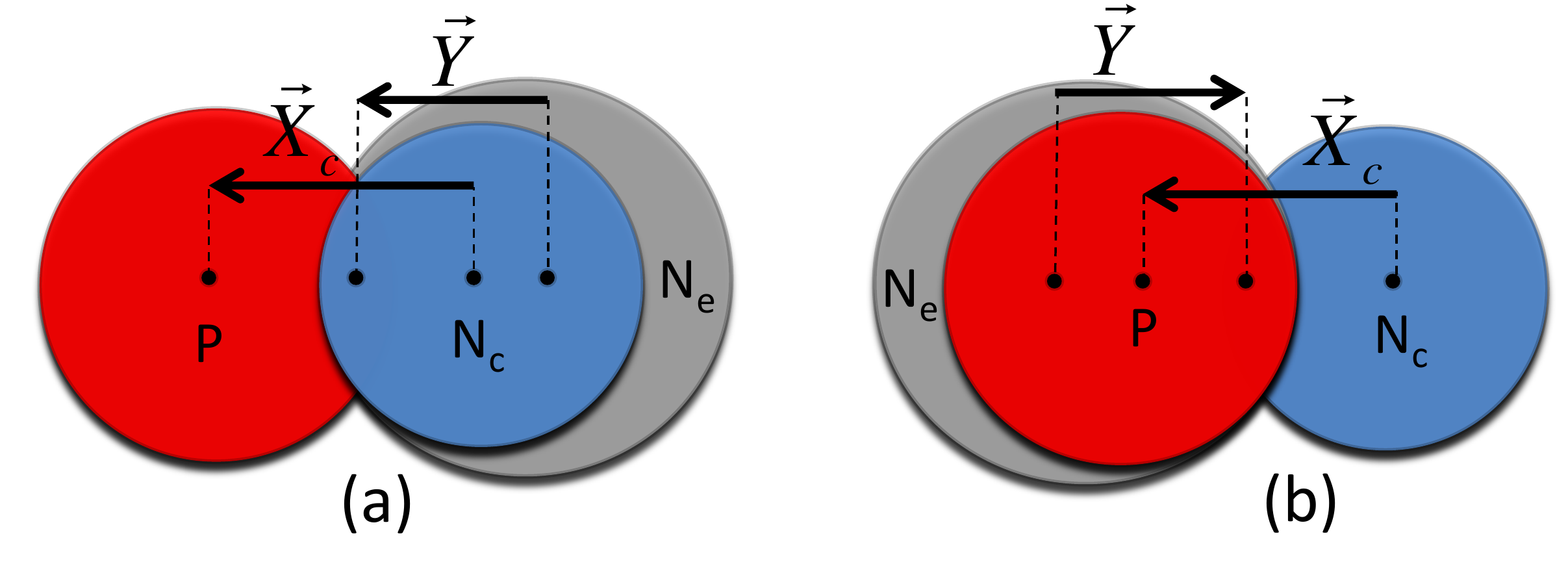}
\end{center}
\caption{(Color online) The coordinates $X_c$ and Y for: (a) GDR (b) PDR, see the text.}
\label{pdrgdr} 
\end{figure}
This picture of GDR and PDR is supported by numerical results obtained within a transport approach based on Vlasov equation \cite{barPRC2012, zhePRC2016}.
 For $^{148}$Sn with $N_e=48$, the time evolution of the distance between the Center of Mass (CM) of protons and the CM of neutrons of the core $\vec{X}_c(t)=\vec{R}_{p}(t)-\vec{R}_{n,c}(t)$, as well as the distance between the CM of the core and the CM of excess neutrons $\vec{Y}(t)=\vec{R}_{c}(t)-\vec{R}_{n,e}(t)$, see Fig. \ref{pdrgdr}, were obtained using an asystiff EOS \cite{barPR2005}. The same collective coordinates can be derived from the schematic model by integrating the position vector over the corresponding density variations defined above. 
 The dipole moment is expressed as $\vec{D}(t)=\frac{ZN_c}{A_c}\vec{X}_c(t)+\frac{ZN_e}{A}\vec{Y}(t)=\vec{D}_c(t)+\vec{D}_Y(t)$. From the imaginary part of its Fourier transform, the $S(E)$ was decomposed as a sum accounting for both degrees of freedom $S(E)=s_0\Im(D(\omega))=s_0(\Im(D_c(\omega))+\Im(D_Y(\omega))) \equiv
S_C(E)+S_Y(E)$ \cite{barPRC2013}, see also \cite{marPRA2008} for an analogous analysis of dipolar oscillations in a coupled boson-fermion system. 
\begin{figure}
\begin{center}
\includegraphics*[scale=0.31]{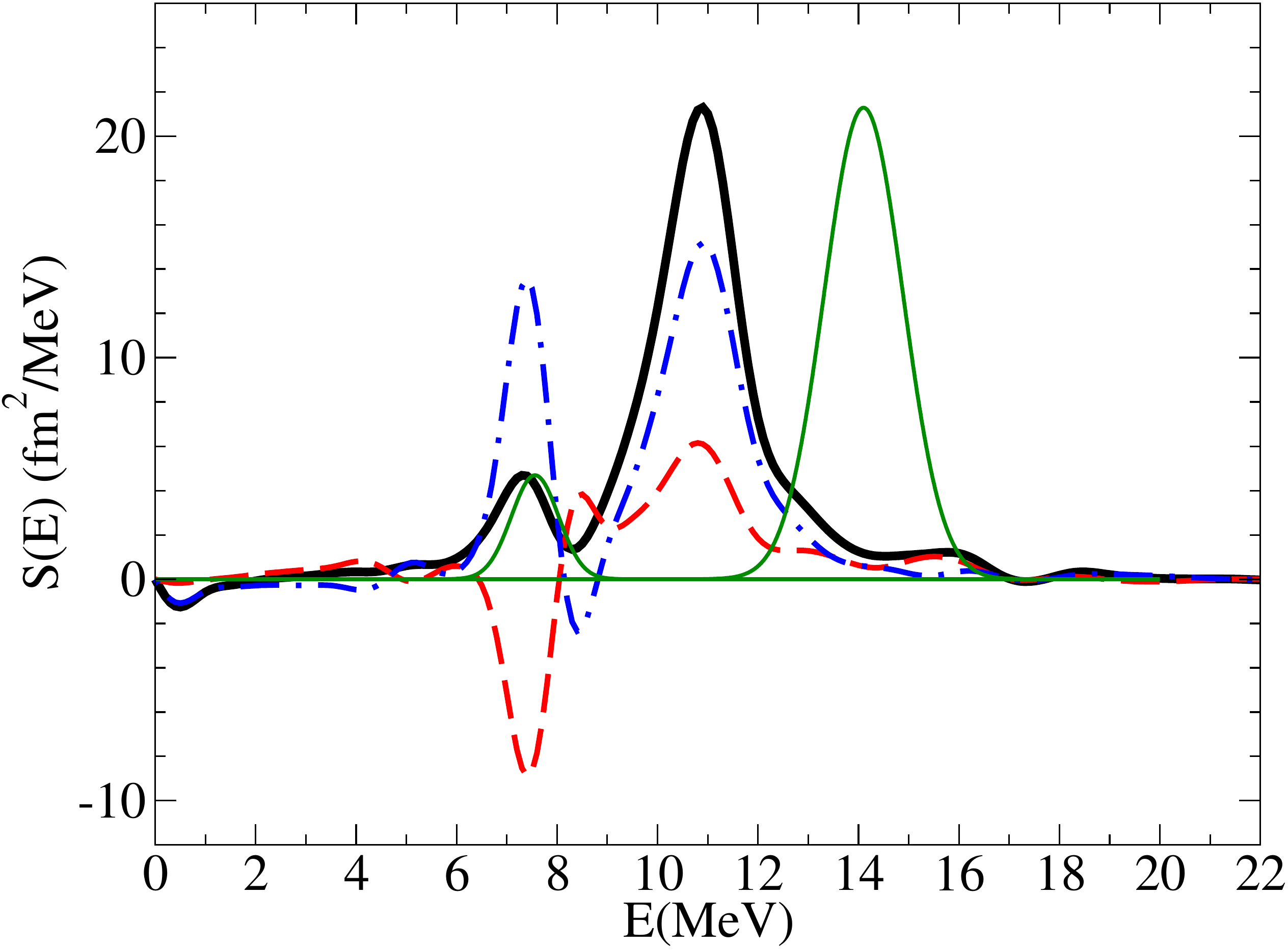}
\end{center}
\caption{(Color online) The total $E1$ strength function $S(E)$ (black (solid) line), the SF's $S_C(E)$ associated to core excitation (blue (dot-dashed) line) and $S_Y(E)$ of the neutrons in excess vibration (red (dashed) line) for $^{148}$Sn, as obtained from Vlasov simulations, see the text. The $S(E)$ corresponding to the schematic model is shown with green (solid) line.}
\label{streyx}
\end{figure}
The EWSR determined from numerical $S(E)$, see Fig. \ref{streyx}, differs by few percentages from the expected TRK sum-rule \cite{barPRC2013}. For comparison, we show in Fig. \ref{streyx} the total $E1$ $S(E)$ as predicted by the schematic model, using a gaussian folding with widths $0.44$ MeV and $1.23$ MeV, for PDR and GDR, respectively (solid green lines). The selected values for the coupling constants $\displaystyle \lambda_2/\lambda_1=0.6, \lambda_3/\lambda_1=0.3$ and $ b_{sym}^{(pot)}(\rho_0)=16$ MeV are similar
to those discussed in \cite{barPRC2015} and determine in an RPA treatment an energy centroid $E_c^{(PDR)}\displaystyle = 7.6$ MeV and an exhausted EWSR fraction $f_{PDR}=\displaystyle 4\%$ when $N_e=48$.  
A good agreement between the two approaches for the PDR centroid and associated EWSR is evidenced, while for GDR the difference in centroid positions is related to a weaker isovector restoring force in finite inhomogeneus systems in Vlasov simulations.
  For vibrations along $z$-axis, if $X_c(\omega)$ and $Y(\omega)$ have the same sign the core neutrons and neutrons in excess oscillate in phase and all neutrons are out of phase with the protons, see Fig. \ref{pdrgdr}. In Fig. \ref{streyx} this is true for the high energy peak of $S(E)$, in agreement with the dynamics of GDR.
When $X_c(\omega)$ and $Y(\omega)$ have opposite signs, the neutrons in excess are in phase with protons and out of phase with respect to the core neutrons, see Fig. \ref{pdrgdr}. This is the case for the low energy peak, around $7.5$ MeV in 
Fig. \ref{streyx}, a picture which agrees with
the dynamics found in the schematic model for PDR, see Eqs. (\ref {densPDRp}), (\ref{densPDRn}). At the PDR peak, $X_c \ne 0$ indicate the core polarization, as is also seen in the schematic model.

\begin{figure}
\begin{center}
\includegraphics*[scale=0.32]{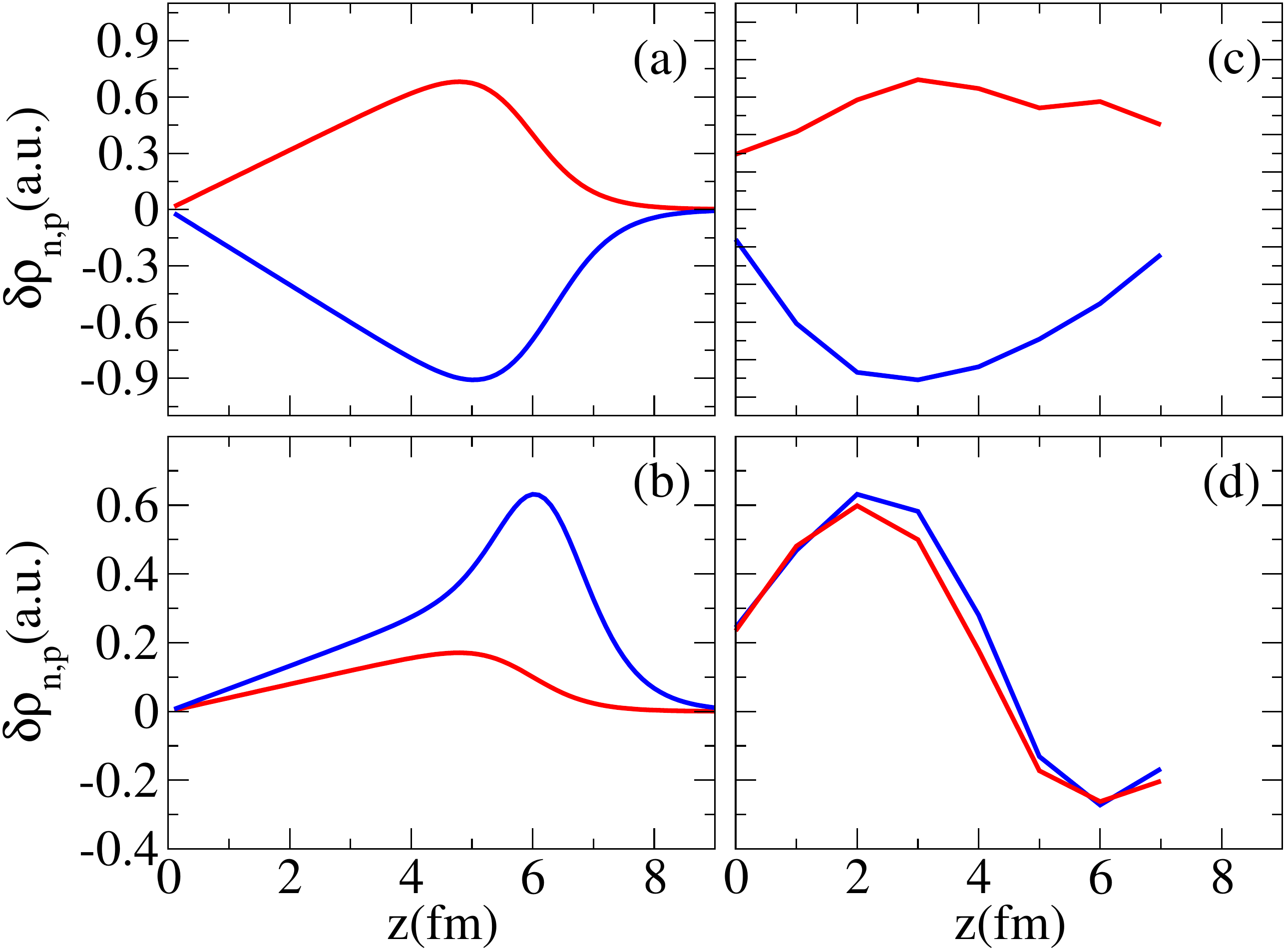}
\end{center}
\caption{(Color online) a) The transition densities for protons (red lines) and neutrons (blue lines) in $^{148}$Sn for: (a) GDR from the schematic model; (b) Same as in (a) but for PDR ; (c) GDR from Vlasov model. (d) Same as in (c) but for PDR. The normalization was obtained from the condition to have the same values for the neutron peaks in (a) and (c) as well as in (b) and (d).}
\label{trdens_pdr}
\end{figure}

The transition densities derived from the schematic model are represented in Fig. \ref{trdens_pdr} (a) and (b) for the two modes, in the case of $^{148}$Sn. A Fermi-type density distribution for neutrons and protons was assumed and
$\rho_{0,n_e}(z)=\rho_{0,n}(z)-\rho_{0,p}(z)$ \cite{isaPRC1992}, obtaining for neutron (proton) radii the values $R_n(R_p)=5.22 (4.80)$ fm.
These exhibit an isovector structure for GDR,  as well as the isoscalar character of PDR. The maximum for the neutron transition densities, close to the surface, is essentially due to the neutron skin. 
We compared these results with the transition densities 
$\delta \rho_{n,p}(E,z)=\int\delta\rho_{n,p}(z,t)\sin(E t/ \hbar)dt$ derived from the densities $\delta\rho_{n,p}(z,t)$ extracted from Vlasov simulations \cite{barPRC2013,barRJP2015,urbPRC2012} for $^{148}$Sn at the energies corresponding to the GDR and PDR peaks of the $S(E)$ shown in Fig. \ref{streyx}. We remark the agreement between the two models, both predicting an isoscalar character of neutron-proton oscillations for PDR. The differences observed can be ascribed to the presence of other excitation components, driven by the isoscalar terms of residual interaction in Vlasov approach not present in the schematic model \cite{zhePRC2016,burPRC2019}.

In conclusion, our study indicates that around the neutron separation energy a collective component of the $E1$ nuclear response becomes visible exhausting some percentage of the EWSR. It was shown that its existence and the isoscalar character are determined by the variation with density of the symmetry energy, which induces a partial decoupling of excess neutrons. For PDR a complex picture of nucleon vibrations resulted, involving both the neutron skin oscillation and isovector core excitation.
In the same energy region states of other physical origin exist, {\it e.g.}, quasiparticles noncollective excitation \cite{tsoPRC2008}, toroidal \cite{repPRC2013,repEPJA2019} or isoscalar dipole \cite{papPRC2014}. The disentanglement of different components is a challenge, theoretically as well as experimentally. The recent possibility to achieve experimentally structured gamma rays which carry orbital angular momentum opens new opportunities to study selectively the nuclear response in the PDR-region by using twisted photons with different angular momenta \cite{taiSCR2017}. The model can be extended to other collective modes where a pygmy-like structure manifests, {\it e.g.}, in nuclear or metallic clusters \cite{palJPB2016}. 

{\it Acknowledgments} V.B. and A.I.N. were supported by a grant from Romanian National Authority for Scientific Research, CNCS-UEFISCDI, Project No. PN-III-P4-ID-PCCF-2016-0164, within PNCDI III and by the project 29 ELI RO financed by the Institute of Atomic Physics. Partial funding by the European Union's Horizon 2020 research 
and innovation program under Grant No. 654002 is acknowledged by M.C..
V.B. and A.I.N. acknowledge discussions with D. Balabanski and D. Stutman on the applications of structured gamma beams. V.B. warmly thanks A. Bonaccorso, M.Di Toro, P. Schuck, H. Wolter  for the fruitful discussions in occasion of the 80th celebration of Massimo Di Toro in Catania, Italy 2019. 
\bibliography{schematic_isoscalar_SHF}
\end{document}